\documentstyle [prl,aps,floats]{revtex}
\input{psfig.tex}
\begin{document}
\def\ba{\begin{eqnarray}}
\def\ea{\end{eqnarray}}
\def\be{\begin{equation}}
\def\ee{\end{equation}}
\def\({\left(}
\def\){\right)}
\def\[{\left[}
\def\]{\right]}
\def\<{\langle}
\def\>{\rangle}
\def\lagrange {{\cal L}}
\def\del {\nabla}
\def\d {\partial}
\def\Tr{{\rm Tr}}
\def\half{{1\over 2}}
\def\fourth{{1\over 8}}
\def\bibi{\bibitem}
\def\S{{\cal S}}
\def\xx{\mbox{\boldmath $x$}}
\newcommand{\labeq}[1] {\label{eq:#1}}
\newcommand{\eqn}[1] {(\ref{eq:#1})}
\newcommand{\labfig}[1] {\label{fig:#1}}
\newcommand{\fig}[1] {\ref{fig:#1}}
\def\phist{\phi_{\rm st}} 
\newcommand{\mb}[1] {\mbox{\boldmath $#1$}}

\title{A Technique for Calculating Quantum Corrections to Solitons} 
\author{Chris
Barnes\thanks{email:barnes@puhep1.princeton.edu},
Neil Turok\thanks{email:N.G.Turok@amtp.cam.ac.uk}} \address{${}^*$Joseph Henry
Laboratory, Princeton University, Princeton, NJ, USA 08540\\
${}^\dagger$DAMTP, Silver St,Cambridge, CB3 9EW, U.K.  }
\date\today 
\maketitle

\begin{abstract}
We present a numerical scheme for calculating the first quantum
corrections to the properties of static solitons.  The technique is
applicable to solitons of arbitrary shape, and may be used in 3+1
dimensions for multiskyrmions or other complicated solitons.  We
report on a test computation in 1+1 dimensions, where we accurately
reproduce the analytical result with minimal numerical effort.
\end{abstract}

\section{Introduction}

The quantum fate of static solitons has long been studied, and
beautiful results have emerged in exactly solvable two dimensional
field theories. In higher dimensions, very little is 
known except in certain very special supersymmetric theories,
or in situations of spherical symmetry.
In many
theories, the quantum properties of the soliton-bearing sector are
interesting, but beyond the reach of current calculations.  One
example of such a theory is the chiral model of nuclear physics, the
solitonic sector of which describes nucleons and nuclei\cite{holst}.
 It is clear
that quantum effects must be included if the theory is to match real
nuclei, but hitherto only a very limited `collective coordinate'
quantization has been possible.  There are many other 3+1 dimensional
field theories with topological solitons, for example those bearing
magnetic monopoles and vortex strings.  It would be very useful to
develop techniques to study quantum corrections to these solitons in
both supersymmetric and non-supersymmetric contexts.

The quantum corrections to a soliton's classical properties may be
expressed as an expansion in a dimensionless parameter, some power of
the coupling constant, multiplied by Planck's constant $\hbar$. At
weak coupling, the corrections are small and may be calculated
perturbatively in this small parameter.  In this paper we outline a
straightforward method for numerically computing the first quantum
corrections, including all fluctuations about a static soliton, to
first order in $\hbar$.  In particular, we present a method for calculating 
the quantum correction to the soliton mass,
although the technique is easily generalized 
to other quantum corrections.  We apply the method to a simple two
dimensional theory where the exact expression is known, and show that
it is reproduced numerically with little computational cost.  We also
discuss the extension of the method to analogous calculations in four
dimensions.

There exist a set of standard techniques for calculating the quantum
corrections to the mass, moments, and other properties of solitons
\cite{rajar}.  The techniques usually used require
knowledge of the phase shifts associated with every possible meson
scattering from the soliton. However in many cases
of interest, for example multi-Skyrmions \cite{multi}, the soliton field is not
spherically symmetric and the necessary information would be extremely
difficult to extract.

In this paper we present a different approach, based upon quantizing
the soliton in a finite box, which provides an infrared regulator and 
reduces the number of modes to be quantised to a discrete set.
Our
method centers upon a formula for the first quantum mass corrections
due to Cahill {\it et al.}\cite{Cahill}, which automatically
removes the worst divergences. The quantum correction to the soliton 
mass is
given as a trace over a complete set of modes.  We compute this trace
by summing first over the lowest normal modes of the soliton, and then
over a plane wave basis up to a finite cutoff.  The contribution of
modes beyond this cutoff is included analytically using a derivative
expansion.  As we shall show, one per cent accuracy in the mass
correction is achievable with very modest computational resources.  We
point toward the application of this technique to more interesting
problems, where it appears feasible to calculate the quantum
corrections for multisolitons in the Skyrme model \cite{skyrme},
 and other 3+1
dimensional theories.

\section{Fluctuations about classical solitons}
\label{flucts}

We would like
to~quantize small fluctuations about some stable, static classical
soliton. The~field's dynamics are determined by a Lagrangian, ${\cal
L}(\phi,\d \phi)$, which gives classical equations of motion
\be
\d_t^2 \phi^a = F^a(\phi,\d\phi,\d^2\phi)\ .  \labeq{motion} 
\ee 
with the soliton ~$\phi^a_{\rm st}(\mb{x})$ as a time-independent solution.
The soliton is a
smooth, localized `lump' in~space, surrounded by
a vacuum of constant field.  
The field $\phi^a(\xx,t)$ may be represented as the sum of the (c-number)
classical soliton,
$\phist(\xx)$, plus a quantum correction $\epsilon(\xx,t)$ which
is an operator obeying canonical commutation relations. We work to first
order in $\hbar$, which amounts to keeping terms up to second order in
$\epsilon(\xx,t)$ in the 
the Hamiltonian. 

In this approximation, the quantum mechanical Hamiltonian is 
\be
 H_T = E_{\rm cl} + :\int d^d x\
{\cal H}_2(\pi^a,\epsilon^a,\d_i\epsilon^a,x):\ , 
\ee
 where $E_{\rm cl} = \int d^d x V(\phist,\d \phist)$ is the classical
soliton mass, and ${\cal H}_2$ 
quadratic in $\epsilon$ and its conjugate momentum $\pi$.
Normal ordering is with respect to
the trivial vacuum normal modes for $\epsilon$, the free mesons, so  
the Hamiltonian
is zero in the soliton-free vacuum.

We may express $\epsilon^a(\xx,t)$ in terms of normal modes 
about the classical soliton. It obeys 
a linear equation
\be
\d_t^2 \epsilon^a = -H^2_{ab} \epsilon^b  \labeq{fluc}
\ee
with $H^2_{ab}$ a positive semi-definite differential operator depending
on the soliton solution. The eigenvalues of $H^2$ are the
frequencies squared $\omega_n^2$ of the normal modes about the soliton. 
We then write
\be
\epsilon^a(\xx,t) = \sum_n 
\[ {e^{-i\omega_n t}\over \sqrt{2 \omega_n}} a_n \epsilon^a_n(\xx) + {\rm c.c.}\]
\ ,
\ee
where $\epsilon^a_n(\xx)$ is the $n$th classical normal mode, and 
$a_n$ is the associated annihilation operator.
In terms of these normal mode operators, the Hamiltonian is
\be
H_T = E_{\rm cl} + :\sum_n \omega_n a_n^\dagger a_n:\ .
\ee
where normal ordering is with respect to the trivial soliton-free 
vacuum. 
The ground state of this Hamiltonian in the presence of a soliton 
can be computed by normal ordering
the second term with respect to the soliton normal modes. 
To do so,
write the operators $a_n^\dagger,a_n$, in terms of
$a^{a\dagger}(\mb{k}),a^a(\mb{k})$, the creation and annihilation
operators for fluctuations about the soliton-free vacuum.
\be
a_n^\dagger = {1\over 2} \sum_{k}\[
\tilde \epsilon^a_n (-\mb{k})a^{a\dagger}(\mb{k})
   \(\sqrt{\omega_n\over\omega_k}  +\sqrt{\omega_k\over\omega_n} \)
+ \tilde \epsilon^a_n (\mb{k})a^a(\mb{k})
   \(\sqrt{\omega_n\over\omega_k}  -\sqrt{\omega_k\over\omega_n} \)
\]\ ,
\ee
where $\omega_k = \sqrt{k^2+m^2}$ are the frequencies of elementary meson
excitations.  Applying this identity, one finds
\ba
 \omega_n a^\dagger_n a_n &=& :\omega_n a^\dagger_n a_n:
+ {1\over 4} \sum_{k}\tilde\epsilon_n^a(\mb{k})
     \tilde\epsilon_n^a (- \mb{k}) 
{(\omega_n-\omega_k )^2\over \omega_k} \\
&=& :\omega_n a^\dagger_n a_n:
+ {1\over4}\left\<n \biggr|{(H-H_0)^2\over H_0} \biggr| n\right\>\ ,
\labeq{ccderiv}
\ea
where $H$ is the single particle Hamiltonian for fluctuations about the
soliton, i.e.,\ the positive square root of $H^2$,
and $H_0$ is the single particle Hamiltonian for fluctuations
about the soliton-free vacuum. 

So, the total Hamiltonian may be written 
\be 
H_T = E_{\rm cl} + \sum_n \omega_n a^\dagger_n a_n +\delta m,
\labeq{cc}
\ee
where $\delta_m$, the quantum correction to the soliton mass, is
\be 
\delta m=  -{1\over 4} {\rm Tr} \[ {(H-H_0)^2 \over H_0}\],
\labeq{regc}
\ee
a result obtained by Cahill, Comtet and Glauber \cite{Cahill}.
The trace is taken over any complete set of states.
In~one dimension, 
\eqn{cc} is finite and no further renormalization is needed.
In~three space dimensions it diverges
logarithmically, and so requires regularization, which may be provided 
by inserting a factor $e^{-\epsilon H_0^2}$. The logarithmic
divergence which results may then be subtracted with 
a local counterterm \cite{BTprog}. Nevertheless, the 
formula \eqn{regc} is still very useful in three dimensions, as it has
removed both the quartic and quadratic divergences. 

To compute the trace (\ref{eq:regc}) we need to~be able to construct
the three ~terms~$\<\alpha| e^{-\epsilon H_0^2}H |\alpha\>$,
$\<\alpha|e^{-\epsilon H_0^2} H_0^{-1}H^2 |\alpha\>$, and~$\<\alpha|
e^{-\epsilon H_0^2}H_0 |\alpha\>$, where the states ~$|\alpha\>$ are
the elements of some basis for field perturbations.  A moments thought
reveals that the last two terms are simple to compute, since we have
an explicit form for the operator $H^2$, and the action $H_0$ on any
perturbation is trivial \eqn{H0times}.  The difficulty lies with
the~first term, the~trace of~$H$ over its positive frequency subspace.
This trouble arises because one does not have the $H$ operator in a
calculationally useful form.

\subsection{The~$H^2$ operator}
In order to calculate the mass correction, \eqn{regc}, we need
the $H^2$ operator, which can be extracted from the classical
perturbation dynamics.  To~second order in
$\epsilon^a$, the perturbation Lagrangian is
\be
{\cal L}(\epsilon ,\d \epsilon)  = \d_t\epsilon^a I_{ab}\d_t\epsilon^b -
\d_i\epsilon^a \(V_{ab;ij}\d_j+W_{ab;i}\)\epsilon^b - 
\epsilon^a M_{ab}\epsilon^b\ ,
\ee
where the tensors $I(\xx),V(\xx),W(\xx),M(\xx)$ are functions of the static
soliton fields. 
The classical time evolution of $\epsilon^a(\xx,t)$ is
\ba
\partial_t^2 
\epsilon^a &=& \[I^{-1}\]^{ab}\(V_{bc;ij}\d_{ij} +\d_iV_{bc;ij}\d_j +
\d_iW_{bc;i} - M_{bc}\)\, \epsilon^c \labeq{lineabove}\\
&=& - H^2\circ \epsilon\ .
\labeq{timeevolution}
\ea
And so the classical equations of motion for the perturbation yield an
operator, $H^2$, whose eigenvalues are the squares of the normal mode 
frequencies of perturbations about the soliton.

\subsection{Low normal modes of the~soliton}
\label{normalmodes}

We have $H^2$; unfortunately its positive square root~$H$ cannot be
easily represented.  So, one cannot calculate all terms in the~trace
\eqn{regc} directly.  However, let us note several points.  First,
the~main~contribution in~\eqn{regc} comes from the~lowest frequency
normal modes of the~soliton; these are the~modes where the~difference
between the~$\< n| H |n\>$ and~$\< n| H_0 | n\>$ are greatest.  Indeed
several authors, e.g. \cite{Scholtz}, \cite{holz}, go so far as to
include {\it only} the~contribution of the~soliton zero modes to
the~trace.  This is not an accurate approximation, however it {\it is}
important to accurately compute the contributions of the lowest
modes. Second, ~$H^2$ shares eigenvectors with~$H$. So, although we
cannot directly construct~$H$, we can find its eigenvectors and
eigenvalues if we can extract them from~$H^2$.  Finally, it is clear
that very short wavelength modes should be accurately described by the
WKB approximation - their contribution to the renormalized mass
correction should be small, local and analytically calculable via a
derivative expansion.

Our strategy is summarized in the following formula:
\ba
{\rm Tr}\[ {\cal O}\] &=&  {\rm Tr}_n \[{\cal O}\] + 
{\rm Tr}_{{}_{k\le k_{\rm max}}}\!
\[(1-P){\cal O}\]
+{\rm Tr}_{{}_{k> k_{\rm max}}}\!\[(1-P){\cal O}\]\ , \labeq{strat} \\ 
{\rm where}\qquad{\cal O}& =& H_0^{-1} (H-H_0)^2 \ . \nonumber
\ea
Here $\{|n\rangle \}$ is a small set of precise, low frequency
normal modes for the soliton, computed numerically as explained below,
and used to take the first trace. The remaining part of the trace is
computed using a simple plane wave basis $e^{i {\bf k.x}}$, corrected
for overcounting using a projection operator $1-P$ with $P= \sum_n
|n\rangle \langle n |$. The contribution from low momentum modes
$k<k_{{\rm max}}$ is computed by straightforward matrix diagonalization. The
final term from modes with $k>k_{{\rm max}}$ is computed analytically via a
local derivative expansion as explained below.

The lowest frequency normal modes of $H^2$ may be extracted using the
real-time method of \cite{us}, or by using $H^2$ to ~drive a diffusion
equation.  The latter method has the advantage that for a limited set
of modes the result converges more rapidly with integration time.  For
a fluctuation $|\epsilon\>$ about the~static soliton, write
\be
\d_\tau|\epsilon\> = - H^2|\epsilon\>\ .
\labeq{diffusion}\ee
Now, create an initial
perturbation~$|\epsilon(0)\>$.  This may be written in~terms of
eigenvectors~$|n\>$ of~$H^2$, with~$ H^2|n\> = \omega_n^2 |n\>$.
\be 
|\epsilon(0)\> = \sum_{n} \epsilon_n|n\>\ ,
\ee
so the~solution of the~diffusion equation \eqn{diffusion} is
\be
|\epsilon(\tau)\> = \sum_n e^{-\omega_n^2 \tau}\epsilon_n |n\>\ .
\ee
Starting in~any state~$|\epsilon\>$, and evolving it forward
according to the~diffusion equation \eqn{diffusion}, the~resulting state 
rapidly becomes dominated by the~lowest frequency mode.

To find the~lowest frequency normal modes of~$H$, one
starts with~an arbitrary
perturbation field $|\epsilon\>$ and evolves it forward according to
the~diffusion equation, until the~resulting field is an acceptably
pure eigenvector of $H^2$.  Call this eigenvector~$|1\>$ and store it.
Next, take another perturbation,~$|\epsilon'\>$, and project~$|1\>$
out of it, and evolve it forward under the~diffusion equation until
one has another eigenvector of~$H^2$.  Orthonormalize this
with~respect to~$|1\>$, and call it~$|2\>$, and store it.  And so on:
in~this way one builds up an archive spanning the~lowest frequency
normal modes of perturbations about the~soliton.  Individual
perturbations in this archive will be a mixture of nearby normal modes
of $H$, since evolving the diffusion equation forward a finite length
of time only separates out modes with substantially different
frequencies.  However, once created, this archive of wavefunctions
will span the space of the the lowest frequency normal modes of $H^2$,
up to a cutoff frequency $\omega_c < \omega_{\max}$, the highest
frequency probed.  From this mixed archive of low normal modes,
one can create a near-perfect archive by diagonalizing the matrix
$\<i|H^2|j\>$, creating a new archive of perturbations, out of linear
combinations of the old ones.  Any new perturbation field $|i'\>$,
will be a near machine-perfect eigenvector of $H^2$ if its frequency
satisfies
\be 
  e^{\({\omega_i'}^2 - \omega_{\max}^2\)T}\ll 1\ ,
\ee
where $T$ is the finite time the diffusion equation evolved forward.
All but the highest handful of modes extracted in this fashion will
be near-perfect; those remaining will be discarded before further
calculation. In the example shown later, we kept only the 9 
lowest frequency modes (including the zero mode) as our set $\{|n\>\} $
of near-perfect low modes.

The contribution of the near-perfect modes is easy to calculate, 
since
\be
H^2|n\> = \omega_n^2|n\>\quad \Longrightarrow\quad H|n\> = \omega_n|n\> 
\labeq{eigdef}\ee
and $H_0$ acts on any perturbation via
\be
H_0 |\epsilon\> = \sum_{k} \sqrt{k^2 + m^2} |\mb{k}\>\<\mb{k}|\epsilon\> \ .
\labeq{H0times}\ee
(Here the~basis~$|\mb{k}\>$ is an abbreviation for all possible plane wave
perturbations of the~field; for several component fields it carries an
internal space index as well as a wavevector.)  And so, in the~basis of normal
modes of $H$, it is straightforward to~get each term in the~trace.

\section{Calculating the~trace over all fluctuations}
\label{fourier_technique}
We wish to calculate the trace (\ref{eq:regc}) over all modes,
however it is not practical to compute all of them as 
above. Instead we proceed by supplementing our near-perfect 
eigenmodes with a trace over plane waves. 

For the set of plane waves we simply construct the matrix
\be
H^2_{kk'} = \<\mb{k}'| H^2 |\mb{k}\>
\ee
for all plane waves $\mb{k},\mb{k}'$ satisfying $k \le k_{\rm
max}$. We then diagonalize this matrix:
\be
H^2_{kk'} = {\cal O}^\top_{kd} \[H^2_D\]_{dd'} {\cal O}_{d'k'}\ ,
\labeq{diag}\ee
where~${\cal O}$ is an orthogonal matrix, and define the 
action of $H$ via 
\be
H_{kk'} = {\cal O}^\top_{kd} \[H^2_D\]_{dd'}^{1\over2} {\cal O}_{d'k'}\ ,
\labeq{diaga}\ee
The~eigenvectors of this matrix give
an approximate basis of normal modes for the~soliton:
\be
 |d\> = \sum_{|k|\le k_{\max}} {\cal O}^\top_{kd}|\mb{k}\>
\labeq{fourierbase}
\ee
and the~eigenvalues~$\[H^2_D\]_{dd}$ yield the~squared frequencies
of these approximate normal modes.   These normal modes will be as 
perfect eigenvectors of $H$ as can be constructed out of the limited
region of Fourier space used. We may now straightforwardly 
compute each of the terms
in the second trace in (\ref{eq:strat}). 
Most simply, if ~$k_{\max}$ is high
enough, the trace will have converged and we are done.

In order for this technique to provide an accurate mass correction,
all terms in the trace, \eqn{regc}, which contribute significantly to
the mass correction must be included.  That is, the subspace of normal
modes spanned by this limited perturbation basis must cover all modes
for which $\<n|(H-H_0)^2/H_0|n\>$ is large.  It is possible to
arrange this, because the finite contibutions from both far infrared
and ultraviolet scattering modes are vanishing as their wavelengths
become greatly larger or smaller than the soliton size.  Let $V_{\rm
sol}$ be the soliton volume scale, and let $k_{\rm sol}$ be a
characteristic soliton wavevector scale, $k_{\rm sol} \sim V_{\rm
sol}^{-1/d}.$ All infrared modes, with wavelength $k<k_{\mbox{\tiny
IR}}\ll k_{\rm sol}$ far away from the soliton give a total
contribution $\sim k_{\mbox{\tiny IR}}^d\(V_{\rm sol}/V_{\rm box}\)$.
After removing analytically calculable corrections (see the next
section), all ultraviolet modes with wavelength $k>k_{\mbox{\tiny
UV}}\gg k_{\rm sol}$ give a total contribution $\sim (1/k_{\mbox{\tiny
UV}})^2$, if the Fourier transform of the soliton field dies away
faster than $1/k^2$ for large $k$.  (The solitons in which we are
interested die exponentially with $k$.) So, the IR and UV cutoffs 
created by working in this truncated Fourier space create only controllable 
errors in the result.

\section{Example: Quantum correction to the mass of the $\phi^4$ kink }
\label{example}
As an example of this technique, we use it to calculate the first
quantum mass correction to a well known example, 
the $\phi^4$ kink in 1+1 dimensions.  This
model has been thoroughly explored; see \cite{rajar} for
example. In the infinite volume limit
the kink solution, and all its 
 perturbation modes are known analytically \cite{exact}.
The calculation 
serves as a useful test 
since our numerical result 
can be compared with the exact analytic expression. 
Of course our technique can be straightforwardly 
be applied to {\it any} soliton bearing 
field theory in 1+1 dimensions.

The Lagrangian for the $\phi^4$ kink is
\be
\lagrange = {1\over 2}\((\d_t\phi)^2 -  (\d_x\phi)^2\) + {1\over 2} m^2 \phi^2 - 
{\lambda\over 4}\phi^4 - {m^4\over 4 \lambda} \ ,
\ee
with the usual double well potential in $\phi$, with minima at $\phi =
\pm m/\sqrt{\lambda}$. The vacuum for this Lagrangian is $\phi$
resting at one or the other minimum.  The kink is a field $\phi(x)$
which starts at one minimum at $x\to-\infty$, and smoothly crosses
over the potential hump at $\phi=0$, to the other minimum as
$x\to\infty$.  Minimizing the energy of such field configurations, one
finds $\phist(x)$ ( $ = m/\sqrt{\lambda}\tanh ( m x/\sqrt{2})$, for
the continuum theory, but the precise form of the function is
unimportant to us.)  Expanding in
small perturbations about the vacuum, $
\phi(x,t) = \phi_{\rm vac} + \epsilon(x,t) 
$, 
we obtain the $H_0^2$ operator:
\be
H_0^2 = -\d_x^2 + (\sqrt{2}m)^2\ .
\labeq{kinkH02}\ee
Similarly, expanding in small perturbations about the kink, we get 
\be
H^2 = -\d_x^2 + \(3\lambda\phist^2(x) - m^2\)\ .
\labeq{kinkH2}\ee
which is the Schrodinger operator for a sech$^2$ potential.
In order to calculate the mass correction, we 
discretise the system and and place it in 
a finite box.  Equations
\eqn{kinkH02},\eqn{kinkH2} are discretised by the simple
replacement $\d^2_x\epsilon \to \(\epsilon_{i-1} - 2 \epsilon_i +
\epsilon_{i+1}\)/\Delta_x^2$.
Boundary conditions are handled by inserting a discontinuous jump of
$2m/\sqrt{\lambda}$ (the distance between the two vacua) at the box
boundary.
This choice of boundary deforms the soliton but
the deformation vanishes exponentially with box size. The perturbations
obey periodic boundary conditions.

\begin{figure}
\centerline{\psfig{file=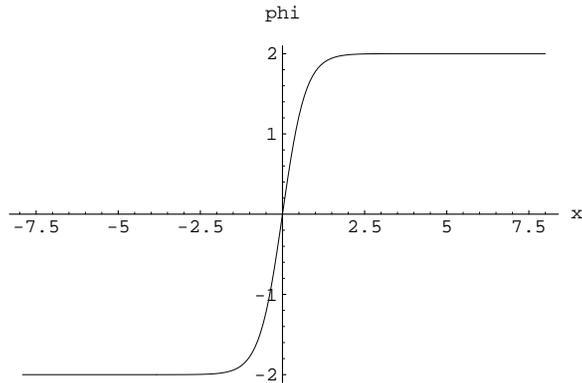,width=3.in}}
\caption{The discretized $\phi^4$ kink. This, and all other figures, 
were calculated with $m=2$, $\lambda=1$, on a 192 point grid, with 
a box size of 16 kink spatial units.}
\labfig{kink}
\end{figure}

We begin by creating a stable soliton
on the grid by relaxing kink-bearing initial conditions to a minimum 
energy configuration.  The relaxed kink is shown in figure \fig{kink}.
The field differs from its continuum value by no more than 
one part in $10^3$ at any point in the box.
Once we have the static soliton, we compute the lowest normal
modes as described above,
solving the  diffusion equation
by the Crank-Nicholson technique
\cite{NR}.  After diagonalization of the $\<i|H^2|j\>$ matrix, 
we obtained an archive of the lowest normal modes of perturbations
about the soliton, each a perfect eigenvector of the discrete $H^2$ operator
to machine accuracy.  A selection of these normal modes is shown in figure
\fig{normal}. 

\begin{figure}
\begin{tabular}{cc}  
\psfig{file=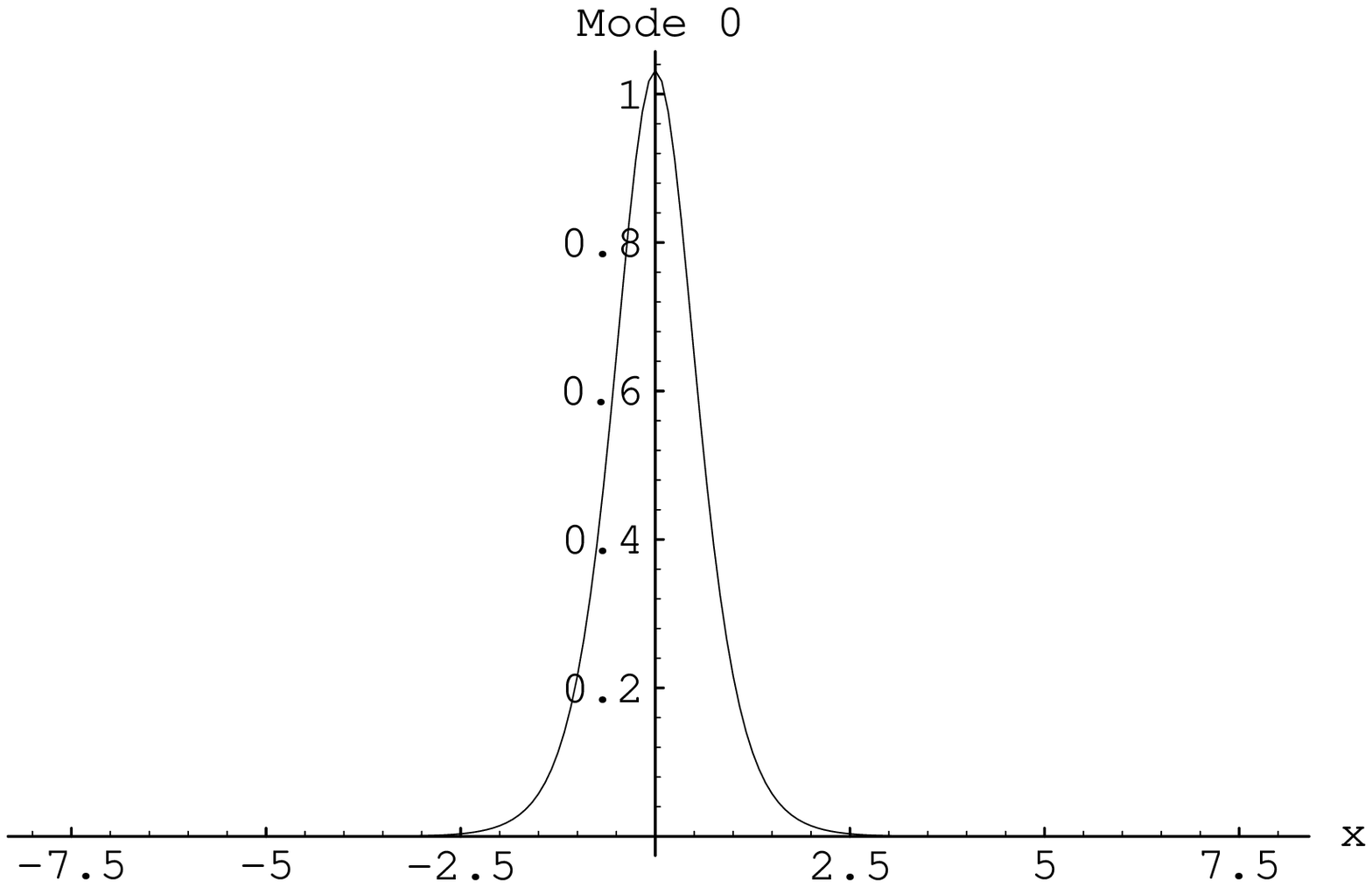,width=3.in}&
\psfig{file=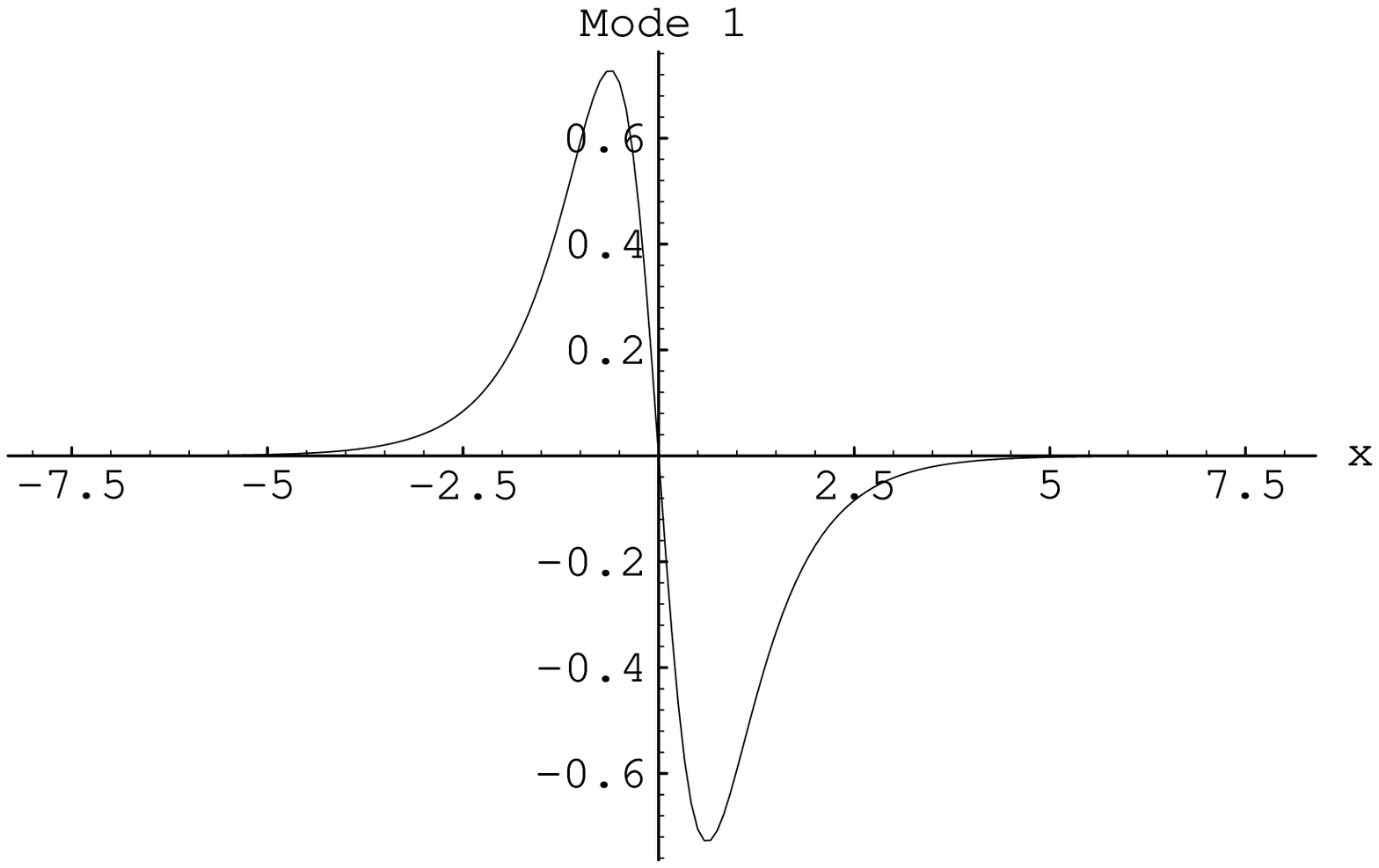,width=3.in}\\
\psfig{file=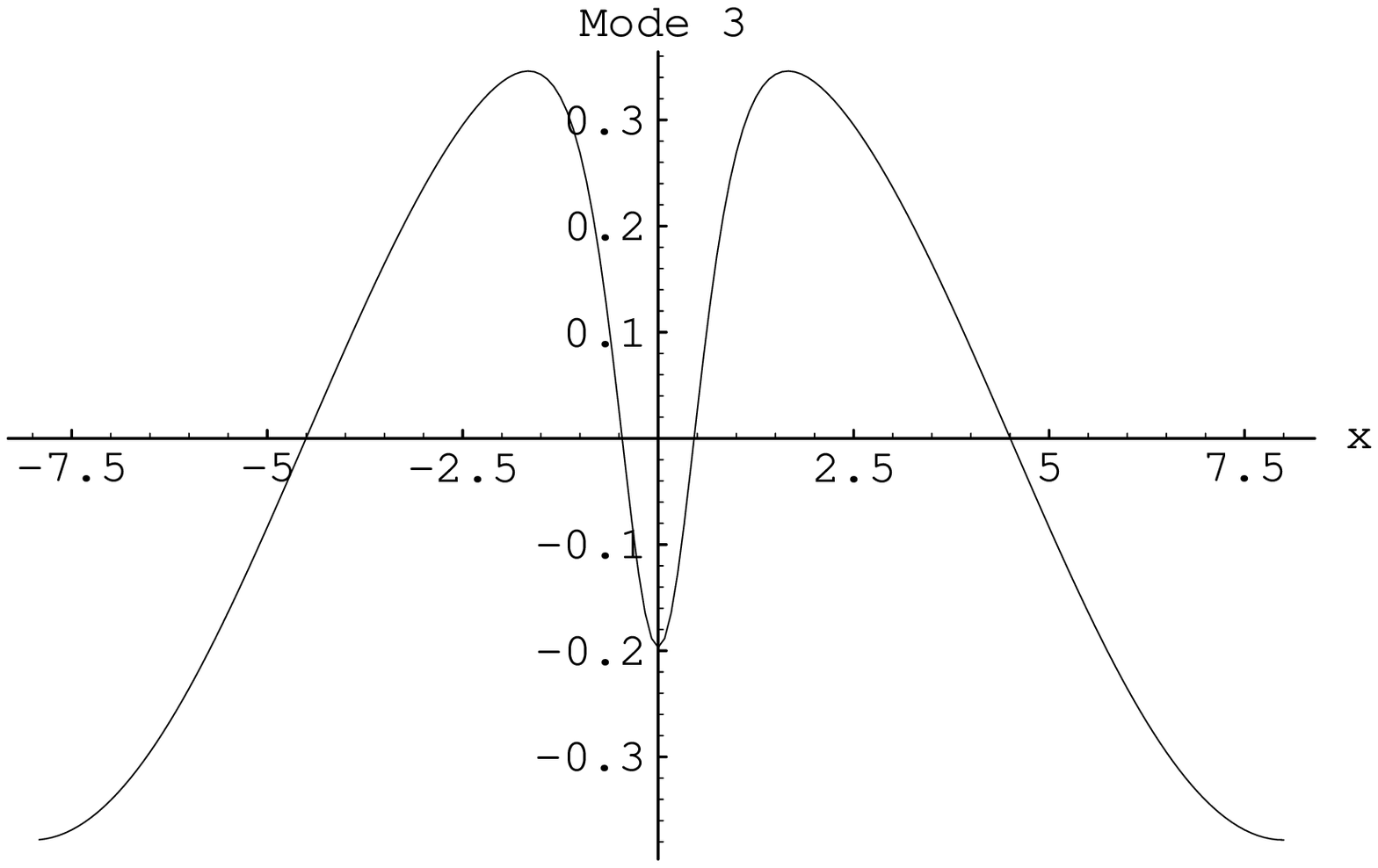,width=3.in}&
\psfig{file=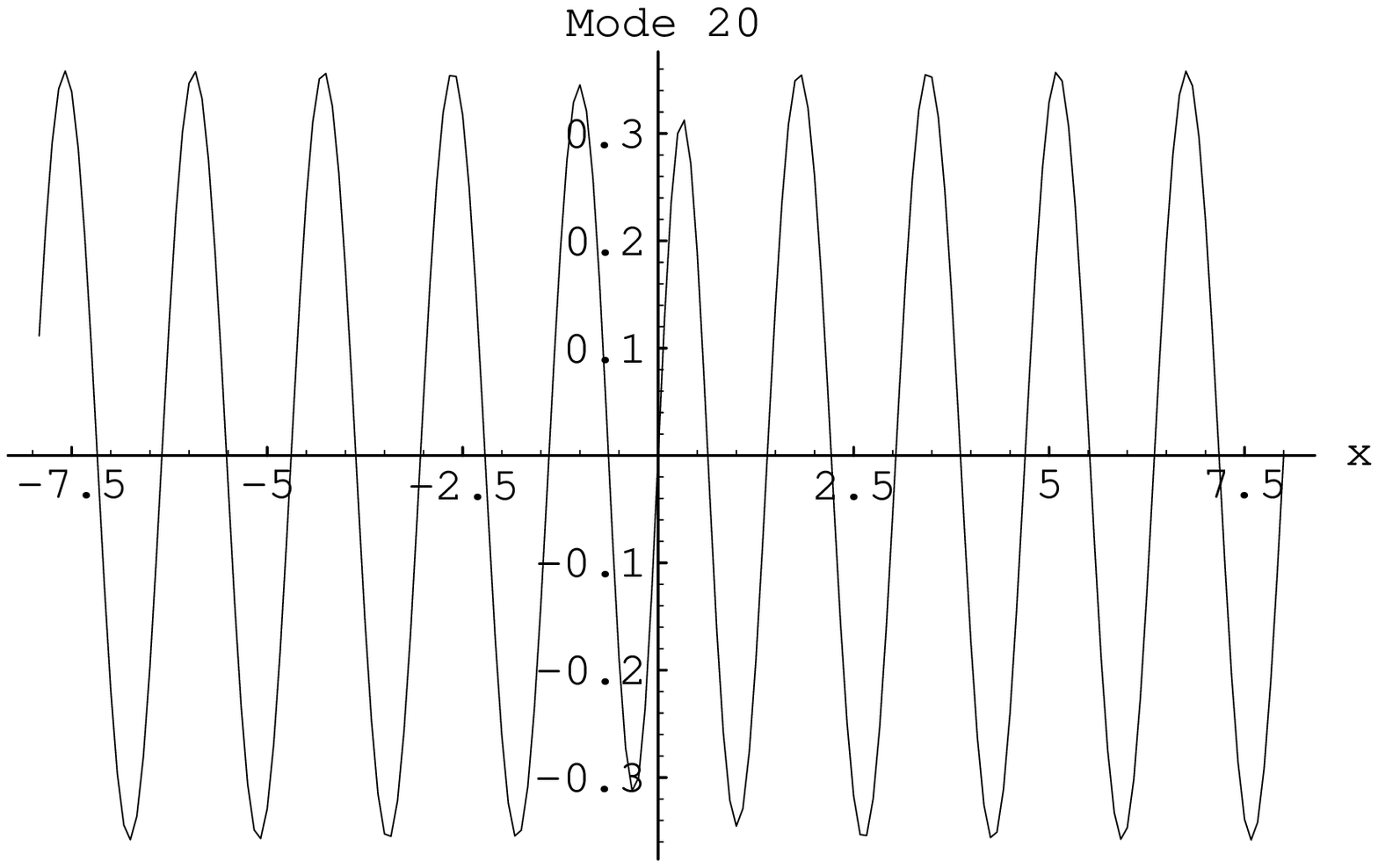,width=3.in}\\
\end{tabular}
\caption{Normal modes of perturbations about the $\phi^4$ kink.  Mode 0
is the zero mode of translations, mode 1 is the bound state (``breathing'')
excitation of the soliton, and the other two are scattering modes. The arbitrary vertical scale has been set by normalizing $\int dx\, \epsilon_n^2 = 1$.
}
\labfig{normal}
\end{figure}

Using these near-exact low
normal modes, together with the set of approximate normal modes
constructed from the Fourier basis as in section
\ref{fourier_technique}, we
calculate terms in the mass correction trace,\eqn{regc}, using
identities \eqn{eigdef},\eqn{H0times}, \eqn{diaga}.
  Including all Fourier modes
in the box, we obtained 
a total mass
correction $\delta m = -0.9468$.  This is to be compared with the
exact result (Rajaraman \cite{rajar}) of $\delta m = m/6\(\sqrt{3/2} -
18/(\pi\sqrt{2})\) = -0.9422$ for these values of parameters.  Thus
this technique generates the right mass correction to about 0.5\%, as
good as could be hoped from a $\sim 200$ point grid. As shown next, 
we do not need to include so many Fourier modes, since the 
high $k$ contribution may be computed analytically.

\bigskip
\section{Analytic Computation of High Momentum Contribution}

The contribution of high $k$ Fourier modes to the mass correction 
can be calculated analytically using an expansion in $k^{-1}$. 
The result can then be used to improve the numerical result
obtained using a limited range in Fourier space. 
The one dimensional theory studied here provides a test case 
for this approach. As we shall see, applying the lowest order correction 
makes the convergence with Fourier mode cutoff very rapid.

As discussed above, the only nontrivial term 
is ${1\over 2} {\rm Tr} H$. We compute this using a 
heat kernel expansion as follows. First we  set 
\be
 H^2= -\partial_x^2 + U(x) +M^2
\labeq{udef}
\ee
with $M= \sqrt{2} m$, the meson mass, and
 $U=\partial^2 V /\partial \phi^2 -M^2$. The potential $U$ defined this
way vanishes at infinity. Next we write
\be
{\rm Tr} H = {1\over \sqrt{\pi}} \int_0^\infty {dt \over \sqrt{t}} \bigl(
-{d \over dt} \bigr)
e^{-t H^2} 
\labeq{heat}
\ee
Performing the trace by summing over a complete set of plane wave states
$e^{i  kx}$ gives 
\be
 {\rm Tr} H = \int {dk \over 2 \pi} \int dx 
\int {dt \over \sqrt{t}} \bigl(-{d \over dt} \bigr)
e^{-t\bigl[-(i k+\partial_x)^2 +U +M^2\bigr]}
\labeq{tr}
\ee
where we used $\partial_x e^{i  kx} = \partial_x +ik$. 
We now separate the exponent into a `free' part $k^2+M^2$ and
an `interaction' part $I=(-2ik\partial_x -\partial_x^2 +U)$.
We then expand $e^{-tI}$ out in powers of $I$, moving derivatives to
the right noting that when a derivative reaches the extreme
right, it gives zero.
Thus we replace $I \rightarrow U(x)$, 
$I^2  \rightarrow U^2-U''-2ik\partial_x U$,  $I^3 \rightarrow 
(-4 k^2) U''$ plus lower order terms in $k$. We then integrate over 
$t$, each power of $t$ giving a factor of $(k^2+M^2)^{-1}$. Keeping 
terms up to $k^{-3}$ we find
\be
{\rm Tr} H = \int {dk \over 2 \pi} \int dx
 \sqrt{k^2+M^2}\(1 +{1\over 2} {U\over {k^2+M^2}}
-{1\over 8} {U^2-U''\over (k^2+M^2)^2} -{1\over 4} {k^2 U''\over (k^2+M^2)^3}
+ ...\)
\labeq{exp}
\ee
where we have ignored terms odd in $k$ since they cancel if we adopt 
a symmetric cutoff $k=|k|$. 

Substituting (\ref{eq:exp}) into the 
Cahill et al. formula (9), the 
first two terms of (\ref{eq:exp}) are cancelled by the last two terms in
(9).
The remainder gives the high $k$ contribution
\be
\delta m_k= -{1\over 8} \int_k^\infty {dk \over 2 \pi} {1\over k^3} 
\int_{-\infty}^{+\infty} dx
(U^2+U'') = -{3M^3 \over 8 \pi \sqrt{2}} k^{-2} + o(k^{-4}).
\labeq{hip}
\ee
We may now use this formula to correct numerical results obtained with
limited coverage in $k$ space.  That is, we compute the 
last term in (\ref{eq:strat}) analytically as above. In this 
calculation we may ignore the term involving the projection 
operator $P$ -- as we shall argue, it vanishes exponentially as
$k_{{\rm max}}$ is increased.  To see this, note that 
the manipulations used in deriving (\ref{eq:hip}) are actually still
valid in the presence of the projector  $P$, but with the introduction of 
a factor $\sum_n|\<k|n\>|^2$ under the $k$ integral. This
involves sum of the squares
 of the Fourier transforms of the mode functions $|n\>$. But
since the latter
are smooth, the factor falls exponentially with $k_{{\rm max}}$.

\begin{figure}
 \centerline{\psfig{file=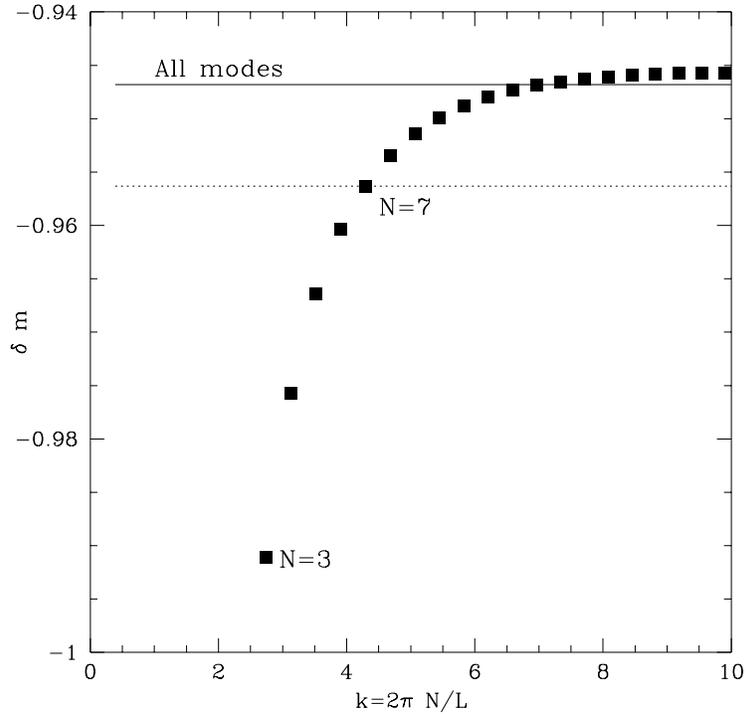,width=4.in}}    
\caption{Convergence of the computation of $\delta m$ 
with cutoff $k=2 \pi N/L$ in a plane wave basis. Solid line 
shows the result including all modes in the box: dashed line
indicates one per cent error, which is reached by $N=7$.  
}
\labfig{conv}
\end{figure}

Figure 3 shows the convergence of the `improved' result as a function of
the cutoff $k$. The horizontal line shows the full result obtained 
by including all modes in the box. The squares show the 
`improved' result obtained for limited numbers of $k$ vectors,
and $N$ labels the cutoff mode number, with $k=2 \pi N/L$.
The `improved' result is within one per cent of the full result 
when we include modes up to N=7, or $k\sim 5$. 

Our technique clearly works very well
for the $\phi^4$ kink in one spatial dimension, and could
be applied with similar ease to any other one dimensional
scalar field theory with kinks.
The convergence of the `improved' formula for $\delta m$ with the
cutoff in $k$ space augurs well for the prospect of performing analogous
calculations in three dimensions. Just as here, the high $k$
contributions can be computed analytically. And if the convergence
is similar, we can expect to obtain accurate results for the
soliton mass corrections using of order $N^3 \sim 10^3$ modes
in $k$ space, which should certainly be feasible.

\section{Conclusion}

We have developed a technique for calculating the first quantum
correction to the mass of a static soliton.  This technique is
implementable for arbitrary stable solitons within a broad class of
Lagrangians.  As an example, we applied this technique to the $\phi^4$
kink, where the result is known analytically: the numerical answer
agreed well with the correct one. We hope to use this technique to
calculate quantum corrections to the masses of more complicated
solitons in three spatial dimensions.

\end{document}